\documentclass[nofootinbib,onecolumn,superscriptaddress,amsmath,notitlepage]{revtex4-1}

\usepackage{graphicx}
\usepackage[colorlinks=true,allcolors=blue]{hyperref}

\begin{document}

\title{Modelling Neutron-Star Ocean Dynamics}

\author{F. Gittins}
\email{f.w.r.gittins@soton.ac.uk}
\affiliation{Mathematical Sciences and STAG Research Centre, University of Southampton, Southampton SO17 1BJ, United Kingdom}
\author{T. Celora}
\affiliation{Mathematical Sciences and STAG Research Centre, University of Southampton, Southampton SO17 1BJ, United Kingdom}
\author{A. Beri}
\affiliation{Department of Physical Sciences, IISER Mohali, Punjab 140306, India}
\author{N. Andersson}
\affiliation{Mathematical Sciences and STAG Research Centre, University of Southampton, Southampton SO17 1BJ, United Kingdom}

\date{\today}

\begin{abstract}
We re-visit the calculation of mode oscillations in the ocean of a rotating neutron star, which may be excited during thermonuclear X-ray bursts. Our present theoretical understanding of ocean modes relies heavily on the traditional approximation, commonly employed in geophysics. The approximation elegantly decouples the radial and angular sectors of the perturbation problem by neglecting the vertical contribution from the Coriolis force. However, as the implicit assumptions underlying it are not as well understood as they ought to be, we examine the traditional approximation and discuss the associated mode solutions. The results demonstrate that, while the approximation may be appropriate in certain contexts,  it may not be accurate for rapidly rotating neutron stars. In addition, using the  shallow-water approximation, we show analytically how the solutions that resemble \textit{r}-modes change their nature in neutron-star oceans to behave like gravity waves. We also outline a simple prescription for lifting Newtonian results in a shallow ocean to general relativity, making the result more realistic.
\end{abstract}

\maketitle

\section{Context}

Precise X-ray timing observations, with instruments like \textit{RXTE} \cite{2006ApJS..163..401J}, \textit{NICER} \cite{2016SPIE.9905E..1HG} and \textit{AstroSat} \cite{2006AdSpR..38.2989A,2022hxga.book...83S}, shed light on neutron-star dynamics and processes related to the star's spin. Relevant phenomena range from waves in the star's ocean (at low densities) to spin variations (like pulsar glitches) and oscillation modes driven unstable by gravitational-wave emission (depending on viscous processes in the high-density core). In principle, the matching of timing observations to reliable theory models should provide constraints on the elusive neutron-star equation of state (both in terms of bulk properties like the mass-radius curve and non-equilibrium transport phenomena leading to viscosity). In this context, quasi-periodic features observed in association with X-ray bursts from accreting neutron stars are tantalising. The frequencies of these burst oscillations are close to the star's spin frequency, although in some cases with an offset of several Hz (in cases where the spin is independently known). The observed frequency  may drift by a few Hz. The intuitive explanation connects the observed features with waves in the shallow neutron-star ocean \cite{2020MNRAS.491.6032C}. This, in turn, suggests a link to  asteroseismology. This paper is motivated by this suggestion. In particular, we want to clarify the assumptions on which current theory models are built and understand to what extent the associated approximations can be seen as robust.

\subsection{Observations}

In order to provide appropriate context, let us first consider the relevant observations. Thermonuclear~(Type~I) X-ray bursts are sudden eruptions due to the unstable burning of hydrogen and helium on the surface of an accreting neutron star in a low-mass X-ray binary \cite{2021ASSL..461..143P,2021ASSL..461..209G,2022ASSL..465..125B}. The fast timing capability of \textit{RXTE} led to the discovery of narrow high-frequency features (mostly in the range $300-600 \, \text{Hz}$) in the power spectrum of these X-ray bursts (see, \textit{e.g.}, Refs.~\cite{1997ApJ...486..355S,1997ApJ...487L..77S}). These features became known as burst oscillations, either found during the rise/decay phase of the burst, or during both phases. The remarkable stability of the asymptotic oscillation frequency indicates that burst oscillations are linked to the neutron-star spin frequency (see, \textit{e.g.}, Ref.~\cite{1998ApJ...503L.147S}). This notion was further supported by the detection of the oscillation frequency in the decay phase of thermonuclear bursts from the accreting millisecond X-ray pulsar SAX~J1808.4$-$3658, which matched well with the known spin frequency ($401 \, \text{Hz}$) of this source \cite{2003Natur.424...42C}. Currently, only a handful (about $20-21$) of neutron-star low-mass X-ray binaries are known to exhibit burst oscillations \cite{watts_anna_l_2021_5513798}. These systems are   also known as nuclear-powered millisecond X-ray pulsars. The observed burst oscillations frequencies are, however, not always perfectly stable. Observations suggest a typical drift of $1-3 \, \text{Hz}$. Conversely, it has been observed that the oscillation frequency does not significantly drift during the burst decay of persistent accreting millisecond X-ray pulsars and nuclear-powered millisecond X-ray pulsars, and typically remains close to the pulsar frequency in these cases. As an example, for XTE~J1814$-$338, the burst oscillation frequency is extremely stable---within $10^{-8} \, \text{Hz}$ of the pulsar spin frequency \cite{2003ApJ...596L..67S}---while for SAX~J1808.4$-$3658, IGR~J17511$-$3057 and IGR~J17498$-$2921,~the former frequency is within a few mHz,~$\sim 0.05 \, \text{Hz}$ and $\sim 0.25 \, \text{Hz}$ of the latter, respectively \cite{2003Natur.424...42C,2010MNRAS.409.1136A,2012MNRAS.422.2351C}. However, \textit{AstroSat} observations of the neutron-star X-ray transient XTE~J1739$-$385 during its 2019--2020 outburst revealed accretion-powered pulsations at $386 \, \text{Hz}$ during very short intervals ($0.5-1 \, \text{s}$) of X-ray flares, making it an intermittent accreting millisecond X-ray pulsar \cite{2023MNRAS.521.5904B}. One of the thermonuclear bursts observed during these observations showed the presence of coherent decay phase burst oscillations at $383 \, \text{Hz}$, indicating a frequency offset of $\sim 3 \, \text{Hz}$ away from the spin frequency of the source. In addition, there have been observations of frequency drifts of $\sim 3.6 \, \text{Hz}$ for 4U~1916$-$053 \cite{2001ApJ...549L..85G} and $\sim  5 \, \text{Hz}$ for MXB~1659$-$298 \cite{2001ApJ...549L..71W}. These frequency drifts---taking place in a few seconds---are significantly larger than the  typical $\le 1 \%$ feature. However, in both cases, the burst oscillation signal drops below the detection threshold in the middle of the burst. Since frequency cannot be tracked continuously throughout the burst, it is not clear whether the observed drift is real \cite{2012ARA&A..50..609W}. One plausible explanation for the observed frequency drift during burst oscillations is the excitation of oscillation modes, the property of which evolves as the ocean settles down after the burst.  Within this explanation, different families of modes (associated with different restoring forces) may lead to a range of observed frequencies. In the following, we aim to shed light on this notion.

\subsection{Theory}

The effort to build theory models that match the observed burst oscillation phenomenology draws heavily on the broader effort to understand neutron-star oscillations, with a significant body of work motivated by issues from, in particular, gravitational-wave astronomy \cite{2019gwa..book.....A}. The equation of state of the low-density neutron-star ocean is naturally different from that of the star's high-density core, but the mathematical machinery required for the problem of ocean waves is pretty much the same as that used for global oscillations. Having said that, the shallow-ocean problems lends itself  to further assumptions. In particular, there is a natural connection with geophysics and work aimed at understanding the Earth's atmosphere and oceans \cite{Eckart1960}. This, in turn, has led to models for neutron-star oceans commonly drawing on the so-called \textit{traditional approximation}. The first study in this direction was conducted by \citet{1996ApJ...460..827B}, who used the approximation to compute low-multipole \textit{g}-modes in the oceans of fairly slowly rotating accreting neutron stars.

Following this, the connection between burst oscillations and ocean modes was first made by  \citet{2004ApJ...600..939H}. He argued---based on the results from \citet{1968RSPTA.262..511L}, which further assume the \textit{shallow-water approximation}---that the observations were best explained by an $|m| = 1$ ``buoyant'' \textit{r}-mode, where $m$ is an integer that characterises the mode's azimuthal dependence. This explanation seemed natural because the \textit{r}-mode solutions occupy a wide band near the equator, resulting in a large variability (required for the oscillations to impact on observations), while these modes have low inertial-frame frequencies (as required by the proximity of the burst oscillation frequency to the star's spin frequency).

Further work on the problem showed that the evolution of the ocean (effectively, cooling after the thermonuclear surface explosion) led to the \textit{r}-mode frequency drifting towards the spin frequency, thus adding further support for the model \cite{2004ApJ...600..914L,2005ApJ...629..438P,2019ApJ...871...61C}. However, Newtonian gravity models suggested that the frequency drift ought to be larger than observed (by perhaps a factor of 2). This issue was addressed by the work of \citet{2020MNRAS.491.6032C}, who extended the traditional approximation calculation to general relativity (drawing on the results from Ref.~\cite{2004MNRAS.351.1349M}). The results from \citet{2020MNRAS.491.6032C} indicate mode frequencies of $\sim 2 \, \text{Hz}$ and frequency drifts of $\sim 2 \, \text{Hz}$. This is more in line with the observations, indicating that relativistic gravity is an important part of any quantitative model (as one would have expected).

At face value, the \textit{r}-mode explanation for the observed quasi-periodic oscillations seems reasonable and the available results are promising. Yet, there is a slight disconnect between the argument and the observations. In particular, it seems difficult to explain systems where the burst oscillation frequency is very close to the spin frequency (as in the case of XTE J1814$-$33, for example). The problem is that, any mode that has a finite frequency in the rotating frame must lead to an offset from the spin frequency in the (observed) inertial frame. This is easy to see from the relationship between the frequencies: 
\begin{equation}
    \sigma = \omega - m \Omega,
    \label{eq:inef}
\end{equation} 
where $\omega$ and $\sigma$ are the mode frequencies as measured by an observer in the frame of the star (rotating with angular spin $\Omega$) and in an inertial frame outside the star, respectively. Clearly, a mode with $m = \pm 1$ and $\omega \ll \Omega$ can lead to $|\sigma|$ being close to $\Omega$, as in the results of \citet{2020MNRAS.491.6032C}, but (and this is a significant but) we cannot have $\sigma \to \Omega$. The only way this can happen is if the ocean dynamics relaxes to geostrophic motion \cite{Pedlosky1987,1989nos..book.....U}, and this is not represented by the \textit{r}-mode solutions.

As things stand, it seems evident that we do not yet have final answers to the questions raised by the observations. The mode explanation is appealing and may well turn out to be correct, but the available results do not allow us to explain all observed systems (some: yes; all: no). The natural next step then is to try to understand to what extent the existing calculations are precise enough to warrant matching against the observations. Given that these models rely (universally) on the traditional approximation it is natural to point the bright light of the inquisitor in this direction. This is our aim with this paper. We want to understand to what extent the traditional approximation is robust for the relatively fast-spinning neutron stars we are interested in. This turns out to be an interesting question in its own right and the answer may (in the extension) help us make progress on the questions that motivated the discussion in the first place. We will, however, not be providing the final answers here.

\section{Ocean oscillations}

In the first instance, let us model the neutron star as a Newtonian, uniformly rotating, perfect fluid. The rotation is assumed to be slow such that the angular frequency $\Omega \ll \Omega_0 \equiv \sqrt{G M / R^3}$ is small, where $M$ is the star's mass and $R$ is its corresponding (non-rotating) radius.%
\footnote{This turns out not to be an overly restrictive constraint, since $\Omega_0 / (2 \pi) \approx 2000 \, \text{Hz}$ for a canonical $M = 1.4 M_\odot$, $R = 10 \, \text{km}$ neutron star, while the fastest known pulsar spin is just above $700 \, \text{Hz}$.}
In this limit, the star is well approximated as a sphere. Moreover, our focus will be on the dynamics of the ocean, representing a thin outer layer, which will be small in mass and contribute little to the star's gravitational potential. Therefore, it is appropriate to adopt the Cowling approximation, neglecting variations in the star's gravitational potential.

We are interested in mode solutions of the form $e^{i (\omega t + m \phi)}$ (\textit{i.e.}, working in the frequency domain and using a Fourier-mode decomposition for the $\phi$-dependence). Hence, in a spherical coordinate basis, the linearised Euler equation in the rotating frame has components 
\begin{subequations}\label{eqs:Euler}
\begin{align}
    - \omega^2 \xi^r - 2 i \omega \Omega r \sin^2 \theta \xi^\phi &= - \frac{g}{\rho} \delta \rho - \frac{1}{\rho} \partial_r \delta p, \label{eq:Euler-r1}\\
    - \omega^2 r \xi^\theta - 2 i \omega \Omega r \cos \theta \sin \theta \xi^\phi &= - \frac{1}{\rho} \frac{\partial_\theta \delta p}{r}, \label{eq:Euler-theta} \\
    - \omega^2 r \sin \theta \xi^\phi + 2 i \omega \Omega (r \cos \theta \xi^\theta + \sin \theta \xi^r) &= - \frac{1}{\rho} \frac{i m \delta p}{r \sin \theta}, \label{eq:Euler-phi1}
\end{align}
\end{subequations}
where $\xi^j$ represents the components of the Lagrangian displacement vector, $\rho$ is the equilibrium mass density, $g$ is the local gravitational acceleration and $\delta \rho$ and $\delta p$ are the Eulerian perturbations of the mass density and pressure, respectively. The mode frequency $\omega$ is as measured by a rotating observer. The perturbations must also satisfy the continuity equation 
\begin{equation}
    \delta \rho = - \frac{1}{r^2} \partial_r (\rho r^2 \xi^r) - \frac{\rho}{\sin \theta} [\partial_\theta (\sin \theta \xi^\theta) + i m \sin \theta \xi^\phi].
    \label{eq:Continuity1}
\end{equation}
To close the system of equations, we require an equation of state. For an adiabatic ocean with frozen composition, it is appropriate to introduce 
\begin{equation}
    \frac{\delta p}{p} = \Gamma_1 \frac{\delta \rho}{\rho} - \Gamma_1 \xi^r \frac{\mathcal{N}^2}{g},
    \label{eq:EquationOfState}
\end{equation}
where $p$ is the equilibrium pressure, $\Gamma_1$ is the adiabatic index and $\mathcal{N}$ is the local Brunt-V{\"a}is{\"a}l{\"a} frequency given by 
\begin{equation}
    \mathcal{N}^2 = - g \left( \frac{d \ln \rho}{dr} - \frac{1}{\Gamma_1} \frac{d \ln p}{dr} \right).
\end{equation}
Although we have made a number of assumptions up to this point, we find ourselves in a situation that plagues much of hydrodynamics. We need to solve a set of coupled partial differential equations~\eqref{eqs:Euler}--\eqref{eq:EquationOfState}. The problem is no less complicated by the presence of rotation. In order to proceed and determine the mode solutions, we need to separate the radial and angular dependencies.

\subsection{The traditional approximation}

The historical approach---with  origins in geophysics---is to use the traditional approximation. The aim of the traditional approximation is simple: We want to effect the separation of the radial dependence from the angular behaviour. This is achieved by disregarding the vertical component of the Coriolis force $\Omega \sin \theta$ in the Euler equation~\eqref{eqs:Euler}. This involves two assumptions. The first is to discard the term with the radial displacement in Eq.~\eqref{eq:Euler-phi1}. This is valid when
\begin{equation}
    |r \xi^\theta| \gg |\xi^r|
    \label{eq:ineq1}
\end{equation}
and should apply for modes that are predominantly horizontal. This is appropriate for some low-frequency oscillations (see Ref.~\cite{2019gwa..book.....A} for an overview of neutron-star oscillations), like \textit{g}- and \textit{r}-modes, but in principle excludes \textit{p}-modes and general inertial modes.%
\footnote{The \textit{r}-modes are a subset of the inertial modes. In general, an inertial mode has comparable radial and horizontal motion, whereas an \textit{r}-mode is unique in that it is axial at leading order (see, \textit{e.g.}, Ref.~\cite{1999ApJ...521..764L}). When a fluid is strongly stratified, the \textit{r}-modes are the only inertial modes that exist.}
Then, the $\phi$-component of the Euler equation becomes 
\begin{equation}
    - \omega^2 r \sin \theta \xi^\phi + 2 i \omega \Omega r \cos \theta \xi^\theta = - \frac{1}{\rho} \frac{i m \delta p}{r \sin \theta}.
    \label{eq:Euler-phi2}
\end{equation}

The second aspect of the approximation removes the angular displacement from the $r$-component of the Euler equation. The motivation for this step is less clear, but one may argue for it based on the expected ocean stratification \cite{1996ApJ...460..827B,1997ApJ...491..839L}. The equation of state~\eqref{eq:EquationOfState} combines with Eq.~\eqref{eq:Euler-r1} to remove $\delta \rho / \rho$, 
\begin{equation}
    - (\omega^2 - \mathcal{N}^2) \xi^r - 2 i \omega \Omega r \sin^2 \theta \xi^\phi = - \frac{1}{\rho} \partial_r \delta p - \frac{g}{\Gamma_1 p} \delta p.
    \label{eq:radeq}
\end{equation}
Based on the first assumption~\eqref{eq:ineq1}, we have the scalings $|\xi^r| \sim h$ and $|r \xi^\theta| \sim |r \sin \theta \xi^\phi| \sim R$, where $h \ll R$ is the depth of the ocean. As indicative values for a neutron-star ocean, we will take $h \sim 10 - 100 \, \text{m}$ \cite{2008LRR....11...10C}. Now, to neglect the Coriolis term in favour of the buoyancy force in Eq.~\eqref{eq:radeq}, we clearly require 
\begin{equation}
    |\mathcal{N}^2| \gg |\omega| \Omega \frac{R}{h}.
    \label{eq:ineq2}
\end{equation}
Due to the relative smallness of $h$ and the assumed low frequencies of the modes we are interested in (recall that the rotating-frame frequency must be small in order to explain the observed phenomenon), this necessitates a combination of a highly stratified layer with an additional restriction on the rotation rate. For example, for accreting neutron stars, Ref.~\cite{1996ApJ...460..827B} argue that this limits the spin to $\Omega / (2 \pi) \lesssim 200 \, \text{Hz}$. However, to retain the first term in Eq.~\eqref{eq:radeq}, we must also have
\begin{equation}
    |\omega| \gg \Omega \frac{R}{h}.
    \label{eq:ineq3}
\end{equation}
Since we have been focusing on modes with low frequencies, this constraint requires the star to be very slowly rotating indeed, certainly beyond what is relevant for the sources we are interested in. In essence, the assumptions associated with the traditional approximation are unlikely to be satisfied for the objects we want to model. Nevertheless, assuming that the two inequalities are satisfied, we arrive at 
\begin{equation}
    - (\omega^2 - \mathcal{N}^2) \xi^r = - \frac{1}{\rho} \partial_r \delta p - \frac{g}{\Gamma_1 p} \delta p.
    \label{eq:Euler-r2}
\end{equation}
Equations~\eqref{eq:Euler-phi2} and \eqref{eq:Euler-r2} are the $\phi$- and $r$-component of the perturbed Euler equation, respectively, in the traditional approximation. The $\theta$-component~\eqref{eq:Euler-theta} is unaltered.

Before we proceed, it is worth commenting on the character of the wave-solutions expected within the traditional approximation. Focusing on short wavelengths, a plane-wave analysis sheds light on this issue (discussed in detail in Appendix~\ref{app:PlaneWave}). The key take-home message is that---contrary to the impression one might get from much of the literature---the traditional approximation is not (strictly) a low-frequency approximation. The equations also support high-frequency sound waves. However, if we insist on consistency and impose Eq.~\eqref{eq:ineq2} [and hence omit the frequency dependence from the radial Euler equation~\eqref{eq:Euler-r2}], then the approximation filters out the sound waves and permits only low-frequency gravity modes (supported by buoyancy). It is worth keeping these observations in mind when making use of the traditional approximation.

In the classic geophysics literature, further approximations are made. The outer layer is modelled as a shallow, uniform-density, incompressible fluid (water, obviously), thereby circumventing (as we discuss later in Sec.~\ref{sec:ShallowWater}) radial derivatives and reducing the problem to one dimension \cite{1898RSPTA.191..139H,Love1913,1964RSPSA.279..446L,1965RSPSA.284...40L,1968RSPTA.262..511L,2021SSRv..217...15Z} . It turns out that, in this context, the traditional approximation becomes necessary in order to ensure that the perturbations conserve angular momentum \cite{Phillips1966,HoltonHakim2013,Vallis2017}. However, this seems to be a pathology of neglecting the radial behaviour, as the radial component of the Euler equation is ignored entirely.

The conceptual issues that arise with the traditional approximation are important, since much of our theoretical understanding of oscillations in neutron-star oceans relies on it \cite{1996ApJ...460..827B,2004ApJ...600..939H,2005ApJ...629..438P,2019ApJ...871...61C,2020MNRAS.491.6032C}. Based on our arguments, it would seem---as also suggested by Ref.~\cite{Eckart1960}---that the traditional approximation is more of an assumption than an approximation. This suggests that care should be taken when applying it to realistic scenarios.

With this caveat in mind, we note that traditional approximation neatly decouples the radial and angular behaviour in the following way. The perturbation equations now show that $\xi^r$, $\delta \rho$ and $\delta p$ share the same polar dependence such that%
\footnote{At this point, it is worth noting that the traditional approximation relies on the Cowling approximation. In relaxing the Cowling approximation, the linearised gravitational potential must be determined from Poisson's equation, which spoils the radial and angular separation that we have arrived at.}
\begin{equation}
    \xi^r = \xi^r(r) \Theta(\mu) e^{i (\omega t + m \phi)}, \quad \delta \rho = \delta \rho(r) \Theta(\mu) e^{i (\omega t + m \phi)}, \quad \delta p = \delta p(r) \Theta(\mu) e^{i (\omega t + m \phi)},
\end{equation}
where $\Theta$ is known as a \textit{Hough function} and $\mu = \cos \theta$ is the latitudinal coordinate. That this separation of variables is possible is a direct consequence of the approximation and is mathematically quite appealing. Given $\Theta$, one is in a position to describe the entire angular behaviour of the perturbations. To see this, the angular components of the Euler equation~\eqref{eq:Euler-theta} and \eqref{eq:Euler-phi2} can be combined to give 
\begin{equation}
    r \xi^\theta = \frac{1}{r \omega^2} \frac{\delta p(r)}{\rho(r)} \hat{\Theta}(\mu) e^{i (\omega t + m \phi)}, \quad r \sin \theta \xi^\phi = \frac{1}{r \omega^2} \frac{\delta p(r)}{\rho(r)} i \tilde{\Theta}(\mu) e^{i (\omega t + m \phi)},
\end{equation}
where 
\begin{subequations}\label{eqs:HoughAngular}
\begin{align}
    \hat{\Theta} &= \frac{1}{1 - (q \mu)^2} \frac{1}{\sqrt{1 - \mu^2}} \left[ - (1 - \mu^2) \frac{d \Theta}{d \mu} + m q \mu \Theta \right], \\
    \tilde{\Theta} &= \frac{1}{1 - (q \mu)^2} \frac{1}{\sqrt{1 - \mu^2}} \left[ - q \mu (1 - \mu^2) \frac{d \Theta}{d \mu} + m \Theta \right].
\end{align}
\end{subequations} 

To determine $\Theta$, we consider the continuity equation~\eqref{eq:Continuity1}, supplemented by Eq.~\eqref{eq:EquationOfState}, to obtain 
\begin{equation}
    \frac{1}{r^2} \frac{d (r^2 \xi^r)}{dr} - \frac{\rho g}{\Gamma_1 p}\xi^r + \frac{\rho}{\Gamma_1 p} \frac{\delta p}{\rho} = \frac{\lambda}{(r \omega)^2} \frac{\delta p}{\rho},
    \label{eq:Continuity2}
\end{equation}
where we have introduced the eigenvalue equation 
\begin{equation}
    \mathcal{L}_q \Theta = - \lambda \Theta
    \label{eq:Laplace}
\end{equation}
with separation constant $\lambda$, operator $\mathcal{L}_q$ defined by 
\begin{equation}
    \mathcal{L}_q \Theta = \frac{d}{d \mu} \left[ \frac{1 - \mu^2}{1 - (q \mu)^2} \frac{d \Theta}{d \mu} \right] - \frac{1}{1 - (q \mu)^2} \left[ \frac{m^2}{1 - \mu^2} + m q \frac{1 + (q \mu)^2}{1 - (q \mu)^2} \right] \Theta
    \label{eq:Operator}
\end{equation}
and spin parameter $q = 2 \Omega / \omega$. Equation~\eqref{eq:Laplace} is known as \textit{Laplace's tidal equation} and fully characterises the angular sector of the perturbation problem. It possesses some noteworthy features.

In the non-rotating limit $q = 0$, Laplace's tidal equation reduces to the associated Legendre equation, where the eigenfunction becomes an associated Legendre polynomial $\Theta = P_l^m$ and the eigenvalue is $\lambda = l (l + 1)$. This reduction is consistent with the familiar fact that a mode of a spherical star may be uniquely described by a single associated Legendre polynomial with a given degree $l$ and order $m$.

All the rotational features are captured by Laplace's tidal equation. However, it is interesting to note that the equation knows nothing about the stratification of the star. This information sits in the radial sector, given by Eqs.~\eqref{eq:Euler-r2} and \eqref{eq:Continuity2}, which is \textit{identical} to that of a non-rotating star when $\lambda = l (l + 1)$ and describes an eigenvalue problem for the mode frequency $\omega$, accompanied by the relevant boundary conditions---zero radial displacement at the ocean base and a vanishing Lagrangian variation of the pressure at the surface. This similarity does allow for a correspondence to the case of spherical symmetry, where useful results can be drawn (see, \textit{e.g.}, Refs.~\cite{1996ApJ...460..827B,1997ApJ...491..839L}).

\section{Solutions of Laplace's tidal equation}

Due to the mathematical usefulness of the traditional approximation, quite some effort has been dedicated to solving Laplace's tidal equation. Notable mentions include Longuet-Higgins's series of seminal papers~\cite{1964RSPSA.279..446L,1965RSPSA.284...40L,1968RSPTA.262..511L}, which discusses the topic extensively. Fortunately, with modern computational techniques and machinery, it is rather straightforward to obtain solutions of Laplace's tidal equation through direct numerical integration \cite{1996ApJ...460..827B,1997ApJ...491..839L}. We will discuss the solutions of Laplace's tidal equation here, paying particular attention to features that do (and, conversely, do not) capture the expected physics.

\begin{figure}[ht]
    \centering
    \includegraphics[scale=0.75]{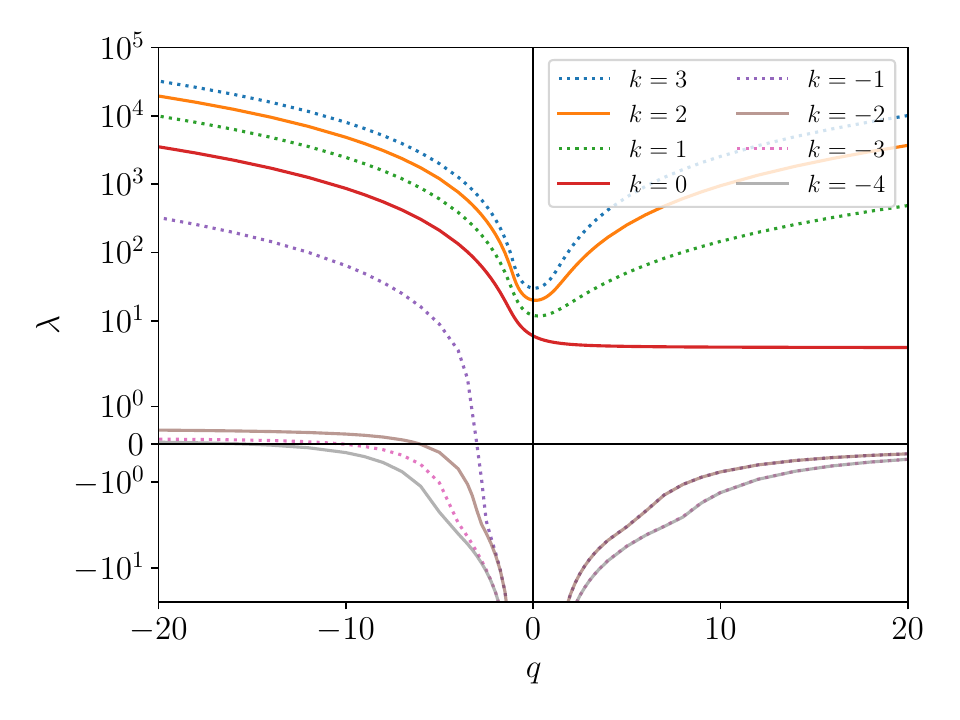}
    \caption{The eigenvalues $\lambda$ of Laplace's tidal equation for $m = -2$ as functions of the spin parameter $q = 2 \Omega / \omega$. Solutions with $q > 0$ are prograde, whereas $q < 0$ correspond to  retrograde perturbations. The $q = 0$ axis corresponds to a non-rotating star. The integer $k$ labels sequences of solutions, where solid and dotted lines indicate even and odd solutions, respectively. The modes from these solutions arise as a consequence of the traditional approximation. The solutions at $q = 0$ correspond to the oscillation modes of a spherical star. With finite $q$, these modes become rotationally modified. The other class of solutions only exists for $|q| > 1$ and can have negative $\lambda$. For $\lambda \geq 0$, this class of solutions capture features of the \textit{r}-modes.}
    \label{fig:lambda}
\end{figure}

Provided a choice of $(m, q)$, one obtains from Eq.~\eqref{eq:Laplace} an infinite set of eigenvalues $\lambda$. As a representative example, we show a selection of eigenvalues for $m = -2$ in Fig.~\ref{fig:lambda}. (Due to the symmetry of Laplace's tidal equation with respect to $m q$, the sign of $q$ can be reversed to give solutions with $m = 2$.) These agree with the results of Refs.~\cite{1968RSPTA.262..511L,1997ApJ...491..839L}. The eigenfunctions $\Theta$ are even or odd, depending on their symmetry about $\mu = 0$ (for more information, see Ref.~\cite{1997ApJ...491..839L}).

As can be seen in Fig.~\ref{fig:lambda}, we can categorise the solutions of Laplace's tidal equation into two classes: (i) perturbations that can exist on the non-rotating star (with label $k \geq 0$) and (ii) those that require rotation ($k < 0$). Since Eq.~\eqref{eq:Laplace} limits to the associated Legendre equation when $q = 0$, we note that class~(i) limit to $\lambda = l (l + 1)$ and can be identified as the familiar polar modes of a spherical star.%
\footnote{In the geophysics literature, these particular solutions are referred to as ``gravity waves'' (see, \textit{e.g.}, Ref.~\cite{1968RSPTA.262..511L}). In fluids that are uniform density and incompressible, it is common in asteroseismology to refer to such perturbations as \textit{f}-modes. The connection between the mode families is, in fact, easy to understand. If we impose a no-penetration condition on the fluid velocity at some fixed depth $h$ then the standard \textit{f}-mode solution (see Ref.~\cite{2019gwa..book.....A} for a pedagogical derivation) limits to that of surface gravity waves when $h\ll R$.}
In general, this includes \textit{f}-, \textit{p}- and (in the presence of stratification) \textit{g}-modes. Although in principle, we would anticipate the high-frequency solutions to be excluded via Eq.~\eqref{eq:ineq1}, the perturbation equations in the traditional approximation do permit such oscillations (as we have shown in Appendix~\ref{app:PlaneWave}). Class~(i) are simply the rotationally modified polar modes.

We follow the convention introduced in Ref.~\cite{1997ApJ...491..839L} and label the sequences with the integer $k$. The integers $k \geq 0$ of class~(i) have been chosen so that, in the limiting case $q = 0$, the solutions correspond to associated Legendre polynomials of degree $l = |m| + k$. Therefore, for $m = -2$, the lowest eigenvalue corresponds to $l = 2$.

Solutions with different signs of $q$ travel in different directions around the star. Modes with $m q < 0$ are prograde---they travel in the same direction as the rotation of the star. On the other hand, modes with $m q > 0$ are retrograde. As we would expect, there is a splitting of the modes when rotation is introduced, with respect to the $q = 0$ axis of Fig.~\ref{fig:lambda}. This splitting is related to the deviation from polar symmetry that arises through the Coriolis force. For a given $|q|$, each mode has a partner that moves in the opposite sense around the star. As is shown in Fig.~\ref{fig:lambda} for $k \geq 0$, the retrograde solutions have larger $\lambda$ than their prograde partners.

We note that the $k = 0$ solution seems to be special among the class~(i) modes, in that it does not reach a minimum $\lambda$ when $q = 0$. For $q > 0$, it asymptotes to a fixed value relatively rapidly. \citet{1968RSPTA.262..511L} described this solution (when $q > 0$) as a prograde Kelvin wave. This seems to be a feature of the assumptions associated with the traditional approximation.

Class~(ii) solutions only arise when $|q| > 1$. Since these solutions require rotation to exist, they belong to the inertial-mode family. One unique aspect of these solutions is that they may acquire negative values of $\lambda$. There has been some discussion of the meaning of such negative eigenvalues. \citet{1968RSPTA.262..511L} proposed that these solutions may be relevant when the system experiences an external force at some fixed frequency. It has also been suggested that they correspond to convective modes that become stabilised by rotation \cite{1997ApJ...491..839L}. Whatever the explanation, the $\lambda < 0$ solutions are of no particular relevance for our discussion. They are simply included in the figure for completeness.

The $\lambda \geq 0$ solutions of class~(ii) are all retrograde. This is a well-known feature of the \textit{r}-modes \cite{1978MNRAS.182..423P,1981A&A....94..126P,1982ApJ...256..717S}. [Recall that we expect general inertial modes to be excluded through the assumption~\eqref{eq:ineq1}.] Furthermore, each of these solutions has the property that when $\lambda = 0$, $m q = (|m| + |k + 1|) (|m| + |k + 1| + 1)$. This is precisely the \textit{r}-mode frequency in the slowly rotating limit, where $l = |m| + |k + 1|$. We show this analytically below. Therefore, the region near to $\lambda = 0$ must be closely related to the slow-rotation approximation.

Finally, there exists a special solution $k = -1$ that has \textit{r}-mode character for small $|q|$, but joins the class~(i) family of solutions when $|q|$ becomes large. This feature was observed by \citet{1968RSPTA.262..511L}. In the slow-rotation limit, this solution corresponds to the $l = |m| = 2$ \textit{r}-mode (with overtones, if the star is stratified). That this \textit{r}-mode joins the rotationally modified polar modes at large $|q|$ is surprising and could also be a feature of the particular assumptions in the traditional approximation.

It is also worth noting that we have, from the outset, assumed the star to be slowly rotating. In theory, this will impact the range of $q$ that Laplace's tidal equation is valid for. By the constraint~\eqref{eq:ineq3}, we arrive at 
\begin{equation}
    |q| \ll \frac{h}{R}.
    \label{eq:ineq4}
\end{equation}
For any shallow layer, where $h \ll R$, this severely limits the values that $q$ can take. This implies that solutions of class~(ii) are unlikely to accurately describe the inertial modes they approximate. This further suggests that care should be taken with interpreting results in the traditional approximation.

\subsection{The shallow-water problem}
\label{sec:ShallowWater}

As we have alluded to above, the first attempts of calculating oscillations in the oceans of rotating bodies neglected the radial dependence entirely, approximating the fluid layer as shallow. In this context, the modes are completely described by Laplace's tidal equation~\eqref{eq:Laplace}. Let us show how this comes out.

In the absence of radial structure, $r = R$ and the $r$-component of the Euler equation~\eqref{eq:Euler-r2} is ignored.%
\footnote{As we ``integrate out'' the radial dependence, we lose any information about stratification in the fluid. In principle, this may still be accounted for by introducing an ``equivalent depth'' (see, \textit{e.g.}, Ref.~\cite{1936RSPSA.156..318T}), but it is not clear to what extent this notion is useful in practice.}
Further, the shallow ocean is treated as a uniform-density, incompressible fluid with $\rho = \text{const}$ and $\delta \rho = 0$. Considering the Lagrangian pressure perturbation at the surface, we then find 
\begin{equation}
    \frac{\delta p}{\rho} = g \xi^r,
\end{equation}
which is assumed to hold throughout the ocean with $g = \text{const}$. This expression may be used to remove $\delta p / \rho$ from the system of equations. It is worth noting that this violates Eq.~\eqref{eq:Euler-r2}, which has been discarded, as well as the boundary condition on $\xi^r$ at the base of the ocean. The shallowness of the layer implies the approximation 
\begin{equation}
    \frac{1}{R^2} \frac{d (R^2 \xi^r)}{dr} \approx \frac{\xi^r}{h}.
\end{equation}
With these assumptions, Eqs.~\eqref{eq:Euler-theta}, \eqref{eq:Continuity1} and \eqref{eq:Euler-phi2} may be expressed in the forms that are found in the classic literature (see, \textit{e.g.}, Refs.~\cite{Love1913,1968RSPTA.262..511L}). Equation~\eqref{eq:Continuity2} readily simplifies to show that 
\begin{equation}
    \lambda = \frac{(R \omega)^2}{g h} = \frac{\epsilon}{q^2},
    \label{eq:Eigenvalues}
\end{equation}
where
\begin{equation}
    \epsilon = \frac{(2 \Omega R)^2}{g h}
    \label{eq:Lamb}
\end{equation}
is the \textit{Lamb parameter} and $\lambda$ is the separation constant we introduced earlier with Laplace's tidal equation~\eqref{eq:Laplace}. Effectively, Eq.~\eqref{eq:Eigenvalues} shows how the two eigenvalues $\lambda$ and $\omega$ become linked in the shallow-water approximation and the problem reduces to a single eigenvalue. However, it is important to keep in mind that the relation~\eqref{eq:Eigenvalues} only holds in the shallow-water approximation.

\begin{figure}[ht]
    \includegraphics[scale=0.43]{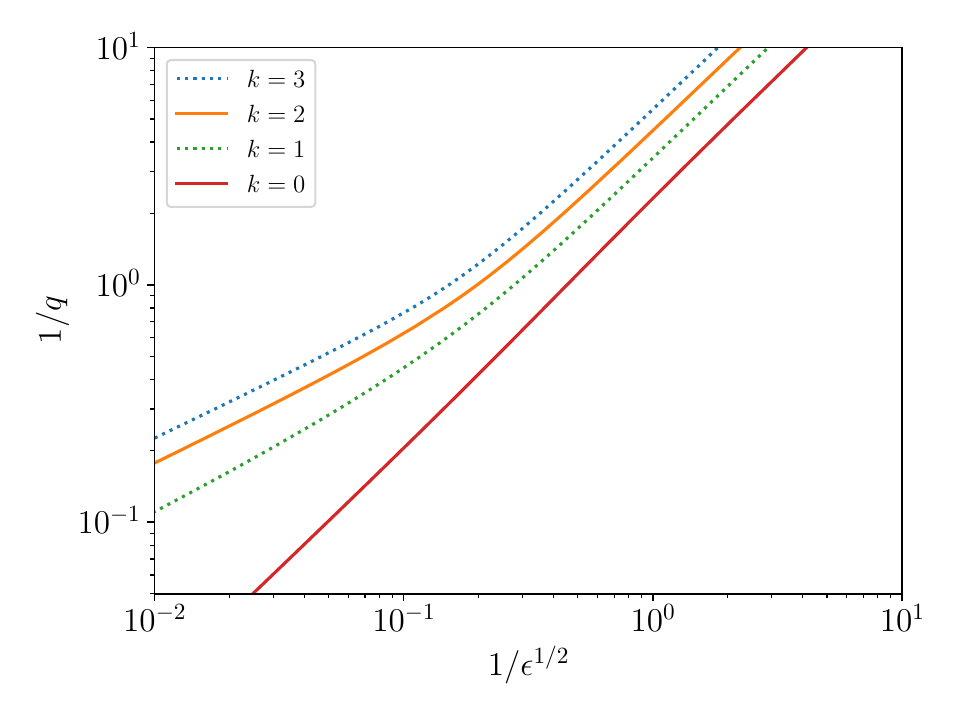}%
    \includegraphics[scale=0.43]{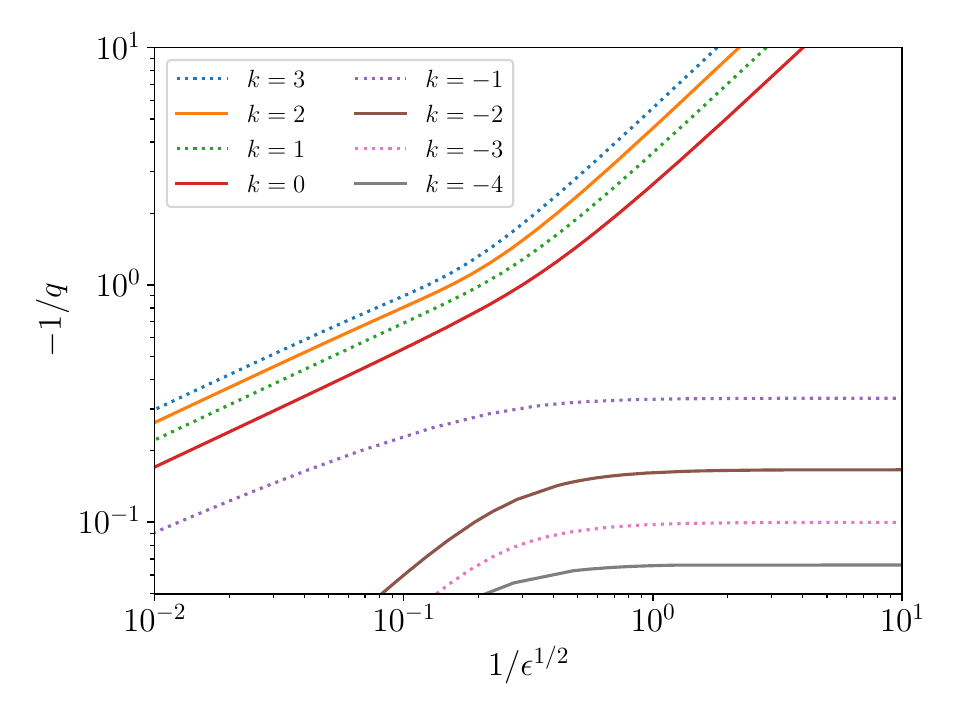}
    \caption{The inverse spin parameter $1 / q = \omega / (2 \Omega)$ against $1 / \epsilon^{1 / 2} = (g h)^{1 / 2} / (2 \Omega R)$ for $m = -2$. The Lamb parameter $\epsilon$ is related to the eigenvalue $\lambda$ by $\epsilon = q^2 \lambda$ when the fluid is assumed to be uniform density, incompressible and shallow. We only show $\epsilon > 0$ solutions. The left panel shows prograde solutions and the right panel shows retrograde solutions. The integer $k$ labels the sequences of solutions, where solid and dotted lines indicate even and odd solutions, respectively.}
    \label{fig:epsilon}
\end{figure}

In the historical shallow-water literature, $\epsilon$ is the favoured eigenvalue. This is accomplished simply by absorbing a factor of $q^2$ into the definition of Eq.~\eqref{eq:Laplace} and solving for $\epsilon$ instead of $\lambda$. Figure~\ref{fig:epsilon} shows the results (for $\lambda > 0$) of Fig.~\ref{fig:lambda} expressed in terms of the Lamb parameter. These are in complete agreement with \citet{1968RSPTA.262..511L} (see his Fig.~3). The left panel shows the prograde perturbations, which only include solutions of class~(i). The right panel shows the retrograde solutions, which include members of class~(i) and (ii). The solutions that asymptote to constant values at large $1 / \epsilon^{1/2}$ approach the expected \textit{r}-mode frequencies.

In the shallow-water approximation, where all the mode features are determined from Laplace's tidal equation, progress can be made analytically, which gives insight into the solutions. (Indeed, this is almost certainly one of the reasons the two approximations, shallow-water and traditional, were adopted in the first place!) With the relation~\eqref{eq:Eigenvalues}, Laplace's tidal equation~\eqref{eq:Laplace} can be brought into the form \cite{1968RSPTA.262..511L}
\begin{equation}
    \left\{ \mathcal{L}_0 + m q + \frac{2 \epsilon \mu}{(m q)^2 - \epsilon (1 - \mu^2)} \left[ (1 - \mu^2) \frac{d}{d \mu} - m q \mu \right] + \epsilon \left( \frac{1}{q^2} - \mu^2 \right) \right\} y = 0,
    \label{eq:LaplaceShallow}
\end{equation}
where $\mathcal{L}_0$ is the operator $\mathcal{L}_q$, defined in Eq.~\eqref{eq:Operator}, when $q = 0$ (which is simply the horizontal Laplacian) and $y = \sqrt{1 - \mu^2} \hat{\Theta}$. Laplace's tidal equation in the form of Eq.~\eqref{eq:LaplaceShallow} is convenient to consider limiting cases.

For small values of $\epsilon / |q|$, Eq.~\eqref{eq:LaplaceShallow} reduces to
\begin{equation}
    (\mathcal{L}_0 + m q - \epsilon \mu^2) y \approx 0.
\end{equation}
We further assume small $\epsilon$ and an immediate solution is found with associated Legendre polynomials for $m q = l (l + 1)$, so
\begin{equation}
    \omega = \frac{2 m \Omega}{l (l + 1)} + \mathcal{O}(\epsilon).
    \label{eq:rmodefreq}
\end{equation}
This is the \textit{r}-mode frequency in the slow-rotation limit. In this regime, the angular dependence of the displacement vector becomes axial [see Eqs.~\eqref{eqs:HoughAngular}]. These asymptotic solutions can be seen in the right panel of Fig.~\ref{fig:epsilon}. However, this regime will not be relevant for fast-rotating neutron stars. This is evident from the Lamb parameter, 
\begin{equation}
    \epsilon = 4 \left( \frac{\Omega}{\Omega_0} \right)^2 \frac{R}{h} \approx 135 \left[ \frac{\Omega / (2 \pi)}{400 \, \text{Hz}} \right]^2 \left( \frac{M}{1.4 M_\odot} \right)^{-1} \left( \frac{R}{10 \, \text{km}} \right)^4 \left( \frac{h}{10 \, \text{m}} \right)^{-1}.
\end{equation}
Clearly, the sources we are interested in lie firmly in the large-$\epsilon$ regime.

Suppose we now examine large $\epsilon / |q|$, then Eq.~\eqref{eq:LaplaceShallow} becomes \cite{1965RSPSA.284...40L} 
\begin{equation}
    \left\{ \mathcal{L}_0 + m q - \frac{2 \mu}{1 - \mu^2} \left[ (1 - \mu^2) \frac{d}{d \mu} - m q \mu \right] + \epsilon \left( \frac{1}{q^2} - \mu^2 \right) \right\} y \approx 0.
\end{equation}
Assuming that we only need to retain the highest derivative to balance the large terms, close to the equator we have 
\begin{equation}
    \left( \frac{d^2}{d x^2} + \frac{A}{\epsilon^{1/2}} - x^2 \right) y \approx 0,
    \label{eq:Hermite}
\end{equation}
where $A = m q + \epsilon / q^2$ and we have changed variables to $x = \epsilon^{1/4} \mu$. This approximation connects directly to Longuet-Higgins's discussion of the asymptotics \cite{1964RSPSA.279..446L,1965RSPSA.284...40L,1968RSPTA.262..511L}. In short, the solutions of Eq.~\eqref{eq:Hermite} are 
\begin{equation}
    y = e^{- x^2 / 2} H_\nu,
\end{equation}
where 
\begin{equation}
    \frac{A}{\epsilon^{1/2}} = 2 \nu + 1
    \label{eq:HermiteIdentification}
\end{equation}
and $H_\nu$ is a Hermite polynomial of integer order $\nu$. Therefore, by Eq.~\eqref{eq:HermiteIdentification}, the mode frequencies are now determined by the cubic equation 
\begin{equation}
    \frac{1}{q^3} - \frac{2 \nu + 1}{\epsilon^{1/2}} \frac{1}{q}  + \frac{m}{\epsilon} = 0.
\end{equation}
For large $\epsilon$, one root is clearly $1 / q \approx m / [(2 \nu + 1) \epsilon^{1/2}]$, leading to
\begin{equation}
    \omega = \frac{m}{2 \nu + 1} \frac{(g h)^{1/2}}{R} + \mathcal{O}(\epsilon^{-1}).
    \label{eq:surfg}
\end{equation}
Notably, this frequency does not depend on the spin of the star. As discussed by \citet{1968RSPTA.262..511L} and shown in Figs.~\ref{fig:lambda} and \ref{fig:epsilon}, all but one ($k = -1$) of the solutions of class~(ii) tend to this frequency.

In terms of the nature of the modes, it follows immediately from Eq.~\eqref{eq:Hermite} that these waves are trapped in the region close to the equator. A standard WKB argument shows that the solution is exponentially small beyond the turning points
\begin{equation}
    x = \pm ( 2\nu +1)^{1/2},
\end{equation}
and it is easy to see that the modes will be equatorially trapped through a Taylor expansion, which leads to
\begin{equation}
    \theta \approx \frac{\pi}{2} \pm \left( \nu + \frac{1}{2} \right)^{1/2} \frac{1}{\epsilon^{1/4}}.
\end{equation}

\citet{2004ApJ...600..939H} used these asymptotic results from \citet{1968RSPTA.262..511L} to argue, based on the structure of the bursts, that the $|m| = 1$ manifestly retrograde solutions of class~(ii) (excluding the $k = -1$ solution that changes character) best explained the observed features. These he termed ``buoyant'' \textit{r}-modes. The terminology seems somewhat confused, however, because the frequency~\eqref{eq:surfg} has the scaling one would expect for a surface gravity wave (and bears no resemblance at all to the \textit{r}-modes, which are linear in the rotation rate). We get a clear hint of what is going on from the right panel in Fig.~\ref{fig:epsilon}. The modes change nature---essentially undergoing an avoided crossing---as the parameters change from the slow-rotation regime (where the solution behaves like the classic \textit{r}-mode) to the fast-rotation regime [where Eq.~\eqref{eq:surfg} applies].

Additional hints that the \textit{r}-modes must change nature follows by repeating the constant-density calculation from Ref.~\cite{2023ApJ...945..139A} with a zero boundary condition imposed at depth $h$. A straightforward calculation leads to the frequency correction (see the original paper for definitions)
\begin{equation}
   \tilde{\omega}_2 \approx - \frac{8m}{(m+1)^4} \frac{1}{(2m+3)} \frac{R}{h}.
\end{equation}
This clearly diverges as $h \ll R$, thus violating the  slow-rotation ordering assumed in the \textit{r}-mode calculation (an issue paid close attention to in Ref.~\cite{2023MNRAS.521.3043G}). In particular, the  $l = |m|$ \textit{r}-mode assumptions require
\begin{equation}
    \tilde{\omega}_2 \ll \omega_0 \left( \frac{\Omega_0}{\Omega} \right)^2 = \frac{2}{m+1}  \left( \frac{\Omega_0}{\Omega} \right)^2, 
\end{equation}
so we need to have 
\begin{equation}
    \frac{h}{R} \gg \frac{4m}{(m+1)^3}  \left( \frac{\Omega}{\Omega_0} \right)^2.
\end{equation}
At the surface of the star this leads to
\begin{equation}
    gh \gg \frac{4m}{(m+1)^3} (\Omega R)^2 \quad \Longrightarrow \quad \epsilon  \ll \frac{(m+1)^3}{m}, 
\end{equation}
so the Lamb parameter cannot be very large. In essence, the $l = |m|$ \textit{r}-modes cannot exist in a shallow neutron-star ocean. This divergence in the frequency correction $\tilde{\omega}_2$ was also observed in numerical calculations we performed following Ref.~\cite{2023MNRAS.521.3043G}, which incorporate rotational shape corrections. This is in accordance with the asymptotic behaviour outlined by Longuet-Higgins \cite{1964RSPSA.279..446L,1965RSPSA.284...40L,1968RSPTA.262..511L}.

Returning to the mode frequency~\eqref{eq:surfg}, let us ask what kind of values we expect. We have
\begin{equation}
    \frac{(g h)^{1/2}}{R} = \Omega_0 \left( \frac{h}{R} \right)^{1/2} \approx 3\times 10^{-2} \Omega_0 \left(\frac{h}{10 \, \text{m}}\right)^{1/2} \left(\frac{R}{10 \, \text{km}}\right)^{-1/2},
\end{equation} 
so, in the inertial frame we get [using Eq.~\eqref{eq:inef}]
\begin{equation}
    \sigma \approx - m \Omega \left[ 1 - \frac{3\times 10^{-2}}{2\nu+1} \frac{\Omega_0 }{\Omega}\left(\frac{h}{10 \, \text{m}}\right)^{1/2} \left(\frac{R}{10 \, \text{km}}\right)^{-1/2} \right].
\end{equation}
In order to be close to the spin frequency we need $|m| = 1$  and the result then  predicts an offset between the spin frequency and the burst oscillation. For the fastest spinning stars, we have $\Omega \approx 0.3\Omega_0$, so it follows that (for $m=1$)
\begin{equation}
    \sigma \approx - \Omega \left[ 1 - \frac{0.1}{2\nu+1}\left(\frac{h}{10 \, \text{m}}\right)^{1/2} \left(\frac{R}{10 \, \text{km}}\right)^{-1/2} \right].
\end{equation} 
If we expect the offset to be 1 \%---in order to connect with typical observations of the bursts (see, \textit{e.g.}, Ref.~\cite{2003Natur.424...42C})---then we need 
\begin{equation}
    0.1\left(\frac{h}{10 \, \text{m}}\right)^{1/2} \left(\frac{R}{10 \, \text{km}}\right)^{-1/2} \approx 0.01 (2\nu+1)
\end{equation}
or
\begin{equation}
    h \approx 10 \left( \frac{2\nu+1}{10} \right)^2 \left(\frac{R}{10 \, \text{km}}\right) \, \text{m},
\end{equation}
so we need $\nu = 4-5$ for $h \approx 10 \, \text{m}$. This does not seem unreasonable.

It is important to note that  we have not considered (since it is not the focus of this work)  the impact of the ocean cooling after the burst. As demonstrated by \citet{2019ApJ...871...61C}, the effect of nuclear reactions in the ocean, accompanied by  varying composition and heat fluxes from the crust, is required to explain the drift of the oscillation modes. In particular, \citet{2020MNRAS.491.6032C} brought this discussion into the arena of general relativity, showing a further reduction in the frequency drifts. Clearly, a realistic calculation of oscillations in neutron-star oceans will need to incorporate all of these features.

\section{Relativistic ocean modes}

Even though we know that accurate theoretical predictions for neutron stars require general relativity to be taken into account, the discussion we have provided so far has been entirely in a Newtonian setting. 
Given this, it is natural  to consider how relativity would enter and modify the results we have discussed. A first answer to this question was provided by \citet{2020MNRAS.491.6032C}, who built on the  relativistic version of the traditional approximation from Ref.~\cite{2004MNRAS.351.1349M}, and calculated oscillations for a cooling ocean following an X-ray burst. The calculation focused on the ``most likely suspects''; retrograde solutions of class~(ii) with $\lambda > 0$ (excluding $k = -1$). \citet{2020MNRAS.491.6032C}  chose $m = 1$ and assumed large (positive) $q$. In the regime of large $|q|$, the eigenvalues of these mode-solutions asymptote to fixed values and become independent of the spin parameter (see Fig.~\ref{fig:lambda} and Ref.~\cite{2005ApJ...629..438P}). Using these eigenvalues---effectively turning the problem into that of a single eigenvalue---\citet{2020MNRAS.491.6032C} solved the relativistic analogues of Eqs.~\eqref{eq:Euler-r2} and \eqref{eq:Continuity2} to obtain the (rotating-frame) frequencies $\omega$. Thus they obtained inertial-frame mode frequencies of $\sigma \sim 2 \, \text{Hz}$ and, by incorporating a model for the ocean cooling, observed drifts of $\sim 2 \, \text{Hz}$, in reasonable agreement with (at least some of) the observations.

Intuitively, we expect three relativistic effects to enter the mode problem for rotating stars. First,  frequencies measured at infinity will involve the gravitational redshift. Second, the rotation will introduce frame-dragging which, in turn, also affects locally measured frequencies. Third, any variation in mass density associated with a wave will generate gravitational waves. Now, for dynamics in the star's low-density ocean, there will not be a significant gravitational-wave contribution. Moreover, as long as the ocean is shallow, we may account for the redshift and the frame-dragging through constant multiplicative factors. This considerably simplifies the problem.

Notably, for modes confined in a thin shell of fluid, the general-relativistic corrections to the frequencies can be quickly obtained following the strategy outlined in Ref.~\cite{2002CQGra..19..191A}. Let us briefly discuss this strategy and apply it to the shallow ocean modes under examination here.%
\footnote{The results obtained here agree with a direct derivation starting from the general-relativistic version of the traditional-approximation equations \cite{2004MNRAS.351.1349M}.} 

For the spacetime of a slowly rotating star, it is generally convenient to work in the coordinate system given by the line element \cite{1967ApJ...150.1005H}
\begin{equation}
    ds^2 = -e^{\nu(r)} dt^2 + e^{\lambda(r)} dr^2 - 2 \varpi(r) r^2 \sin^2\theta\, dt\,d\phi + r^2 (d\theta^2 + \sin^2\theta\, d\phi^2).
\end{equation}
As we are interested in ocean modes, we specialise to a thin shell on the star's surface $r=R$.
Because the frame-dragging is small $\varpi \sim \mathcal{O}(\Omega / \Omega_0)$, the line element on a thin shell at the surface of the star can also be written as
\begin{equation}
      ds^2 = -e^{\nu_R} dt^2 + R^2 d\theta^2 + R^2 \sin^2\theta (d\phi -\varpi_Rd t)^2,
      \label{eq:ThinShellSlowRotMetric}
\end{equation}
where the metric potentials $\nu_R$ and $\varpi_R$ are approximated by their values at the surface---and it is worth noting that this is consistent with Ref.~\cite{2020MNRAS.491.6032C}. It is then easy to see that, via the simple coordinate transformation 
\begin{equation}
    t \to \tilde{t} = e^{\nu_R/2} t, \qquad \theta \to \tilde{\theta} =\theta, \qquad \phi \to \tilde{\phi} = \phi - \varpi_R t,
    \label{eq:AbramoviczTransf}
\end{equation}
the line element~\eqref{eq:ThinShellSlowRotMetric} takes the Minkowski form (on a thin shell and in spherical coordinates) 
\begin{equation}\label{eq:MinkShellMetric}
    ds^2 = - d\tilde{t}^2  + R^2 ( d\tilde{\theta}^2 + \sin^2\tilde{\theta} \, d\tilde{\phi}^2 ).
\end{equation}
Hence, we have transformed the coordinates to a (flat) Lorentz frame with $r = R = \text{const}$. This is a convenience of the shallowness of the layer we are considering. Moreover, the slow-rotation assumption implies that the fluid elements in the star move at non-relativistic velocities \cite{1967ApJ...150.1005H}. The upshot to this is that we can take Newtonian results (provided they are confined to a thin shell, the case of interest here) and use the transformation~\eqref{eq:AbramoviczTransf} to derive the corresponding result in relativity.

Let us show how we can take advantage of this in practice. 
The first thing to note is that we first need to convert from rotating frame to inertial-frame frequencies as the line element~\eqref{eq:MinkShellMetric} is of Minkowskian form. Taking the shallow-water result as a starting point, this means that the inertial-frame frequency is [by Eq.~\eqref{eq:surfg}]
\begin{equation}
    \sigma_\text{N} = -m \Omega_\text{N} + \frac{m}{2\nu +1} \frac{(g h)^{1/2}}{R},
    \label{eq:NonRotFreq}
\end{equation}
where we have explicitly introduced the ``N'' subscript to emphasise that the quantities are calculated assuming Newtonian gravity. Next, using the fact that the phase $\omega t + m \phi$ is an invariant quantity, together with the transformation~\eqref{eq:AbramoviczTransf}, we readily obtain the useful correspondence 
\begin{equation}
    \sigma_\text{GR} = e^{\nu_R /2} \sigma_\text{N} - m \varpi_R,
    \label{eq:FreqTransf}
\end{equation}
where, as before, $\nu_R$ and $\varpi_R$ are the redshift factor and the frame-dragging evaluated at the surface of the star, respectively, and we have the ``GR'' label to denote relativistic quantities. The last step is to note that also the angular velocity as measured at infinity is going to change from Newtonian theory to relativity. Noting that this is defined as $d \phi/ d t$ and using the transformation~\eqref{eq:AbramoviczTransf}, we obtain 
\begin{equation}
    \sigma_\text{GR} = e^{\nu_R /2} \Omega_\text{N} + \varpi_R.
    \label{eq:AngVelTransf}
\end{equation}
Finally, using Eqs.~\eqref{eq:NonRotFreq}--\eqref{eq:AngVelTransf}, we arrive at the final result 
\begin{equation}
    \sigma_\text{GR} = -m \Omega_\text{GR} + \frac{m}{2\nu+1 } e ^{\nu_R/2} \frac{(g h)^{1/2}}{R}.
\end{equation}
We see that the frame-dragging has no effect whatsoever, while the redshift factor enters explicitly. This result is consistent with what we might have expected from the beginning, as noted by, for example, Ref.~\cite{2005ApJ...629..438P}. It is important to stress, however, that this is due to the fact that the rotating frame frequency in the large-$\epsilon$ limit does not depend on the spin frequency. It is not true in general. As an explicit demonstration, we may follow the same steps, this time instead starting from Eq.~\eqref{eq:rmodefreq}---although we should (obviously)  keep in mind that this result follows in the small-$\epsilon$ limit, which does not seem to be appropriate for the systems we are interested in here. Anyway, in this case we arrive at 
\begin{equation}
    \sigma_\text{GR} = - m\Omega_\text{GR} + \frac{2m}{l(l+1)}\left(\Omega_\text{GR} - \varpi_R\right).
\end{equation}
Notably, the frame dragging now enters the relativistic expression explicitly, while the redshift factor does not. 

Finally, let us estimate the magnitude of the relativistic effects in the large-$\epsilon$ limit. For a fiducial neutron star of $M = 1.4 M_\odot$ and $R = 10 \, \text{km}$, we have
\begin{equation}
    e^{\nu_R/2 } = \left(1 - \frac{2 G M}{c^2 R}\right)^{1/4} \approx 0.87,
\end{equation}
so that, in order for the frequency  offset to be approximately $1\%$ we need 
\begin{equation}
    h \approx 10 \left(\frac{2\nu+1}{8.7}\right)^2\left(\frac{R}{10 \, \text{km}}\right) \, \text{m}.
\end{equation}
Not surprisingly, accounting for general-relativistic corrections changes the Newtonian prediction by $30 \%$, in good agreement with the numerical results from 
\citet{2020MNRAS.491.6032C}.

\section{Concluding remarks}

The traditional approximation finds its roots in the classic geophysics literature and has also been used to calculate the dynamics of neutron-star oceans. However, the conditions under which the approximation holds are not particularly well understood. In an attempt to demystify it, we examined the implicit assumptions that underlie this approximation. For the low-frequency modes that we are interested in (in order to describe the small offsets in some X-ray detections), we found that the traditional approximation necessitates the neutron star to be rotating far slower than the observed systems. Thus, we suggest that the approximation may be inaccurate for the low-frequency oscillations of neutron-star oceans.

The central appeal of the traditional approximation is that it decouples the radial and angular dependencies of the perturbations. The angular sector of the perturbation problem becomes solely described by Laplace's tidal equation, associated with the eigenvalue $\lambda$, while the radial sector holds information about the stratification of the fluid and corresponds to the mode frequency $\omega$. We discussed the eigenvalues of Laplace's tidal equation, which belong to two classes. Class~(i) are the rotationally modified polar modes, which exist on the non-rotating star, and class~(ii) include particular inertial modes, which limit to the \textit{r}-modes in the slowly rotating regime. We noted the interesting behaviour of two solutions. (1) The $k = 0$ perturbation, previously known as a prograde Kelvin wave for $q > 0$, asymptotes rapidly to a minimum, which is contrast with other class~(i) solutions. (2) At small $\lambda$, the $k = -1$ solution corresponds to the $l = |m|$ \textit{r}-mode. However, as $|q|$ becomes large, this perturbation joins the class~(i) solutions.

To relate to the classic literature, we discussed the shallow-water approximation, where the layer is assumed thin, uniform density and incompressible. Under these assumptions, the mode problem reduces to one eigenvalue, the Lamb parameter $\epsilon$, and is solely angular. We examined the large-$\epsilon$ limit, appropriate for neutron-star oceans. We showed how the character of the class~(ii) solutions (excluding the special $k = -1$ perturbation) change nature in this regime and no longer behave like inertial modes. This is supported by the standard mode calculation, which shows that the classic \textit{r}-modes cannot exist in a shallow neutron-star ocean.

Finally, taking a step towards a more realistic setting for neutron stars, we demonstrated a simple prescription for lifting Newtonian results in a shallow layer to general relativity. While being a useful procedure for straightforwardly generalising Newtonian calculations, it also illustrates the importance of relativity. We found that relativistic corrections adjust calculations in Newtonian gravity by $30 \%$.

Moving forward, an obvious improvement would be to ensure consistency within the traditional approximation, thereby making it an appropriate low-frequency filter. Fortunately, the resolution is basically trivial: One simply neglects the mode frequency in the radial component of the Euler equation~\eqref{eq:Euler-r2}. This will spoil the precise correspondence between the traditional approximation and the perturbation equations in spherical symmetry, but this seems like a small price to pay. To what extent this impacts on the numerical mode-solutions remains to be seen.

Additionally, it is worth recalling that one of the assumptions within the traditional approximation is to assume the fluid is rotating sufficiently slowly such that it may safely be approximated as spherical in shape. For bodies like our Earth, this is a perfectly reasonable constraint. However, we know that neutron stars can rotate very rapidly indeed. Along this vein, there has been work developing the perturbation equations in the traditional approximation to include the centrifugal deformation of the star \cite{2019A&A...631A..26M,2020MNRAS.496.2098V}. Other aspects of neutron-star physics will inevitably be important to incorporate into any detailed model of these systems. This includes (but is not limited to) the crust and magnetic field. However, in both of these contexts, we will need to depart from the traditional approximation. It would, of course, be interesting to progress these ideas further.

The issues we have discussed were important to raise, since our quantitative understanding of the natural oscillation modes of neutron-star oceans relies almost entirely on the traditional approximation. The interest in this direction has largely stemmed from using these mode solutions to explain the timing phenomenology of X-ray bursts on accreting neutron stars. Ultimately, we have shown that one should be careful in applying the traditional approximation to this context. We must also re-emphasise that a number of the observations exhibit bursts at the same frequency as the star's spin and a mode with a finite rotating-frame frequency will \textit{necessarily} result in an offset. Given this tension, modes can describe some observations, but certainly not all.

\begin{acknowledgments}
FG and NA gratefully acknowledge support from STFC via grant number ST/V000551/1. AB is funded by an INSPIRE Faculty grant (DST/INSPIRE/04/2018/001265) by the Department of Science and Technology, Govt. of India. She is grateful to the Royal Society, UK and also acknowledges the financial support of ISRO under \textit{AstroSat} archival Data utilization program (No. DS-2B-13013(2)/4/2019-Sec. 2). The authors are grateful to A.~L. Watts for helpful comments.
\end{acknowledgments}

\appendix

\section{Plane-wave analysis}
\label{app:PlaneWave}

In order to understand the implications of the traditional approximation, let us consider a simple plane-wave analysis of the radial problem. We start from the $r$-component of the Euler equation~\eqref{eq:Euler-r2} and the continuity equation~\eqref{eq:Continuity2} in the following forms: 
\begin{align}
    (\mathcal{N}^2-\omega^2) \xi^r + \frac{1}{\rho} \left[ \partial_r \delta p  + \frac{g}{c_\text{s}^2} \left( 1 -  \frac{\mathcal N^2 c_\text{s}^2}{g^2} \right) \delta p \right] &= 0, \label{eq:first}\\
    \frac{1}{c_\text{s}^2} \delta p - \frac{\mathcal{N}^2}{g^2} ( \delta p - \rho g \xi^r )  + \frac{1}{r^2} \partial_r (\rho r^2 \xi^r) - \frac{\lambda(q)}{r^2\omega^2} \delta p &= 0, \label{eq:second}
\end{align}
where the background speed of sound $c_\text{s}$ is defined as
\begin{equation}
    c_\text{s}^2 = \frac{dp}{d\rho}
\end{equation}
and it it important to keep in mind that the separation ``constant'' $\lambda$ really depends on both $\omega$ and $\Omega$ (through the spin parameter $q = 2 \Omega / \omega$). Rewrite the first equation~\eqref{eq:first} using an integrating factor
\begin{equation}
    \partial_r \bar{p} - \frac{\mathcal{N}^2}{g} \bar{p} = (\omega^2 - \mathcal{N}^2) \rho \xi^r \exp\left(\int \frac{g}{c_\text{s}^2} dr\right),
\end{equation}
where 
\begin{equation}
    \bar{p} = \delta p \exp\left(\int \frac{g}{c_\text{s}^2} dr\right),
\end{equation}
while the second equation~\eqref{eq:second} becomes
\begin{equation}
    \frac{1}{r^2} \partial_r (\rho r^2 \xi^r) + \frac{\rho \mathcal{N}^2}{g} \xi^r  = \left[  \frac{\lambda(q)}{r^2\omega^2} - \frac{1}{c_\text{s}^2} + \frac{\mathcal{N}^2}{g^2} \right] \bar{p} e^{-\int g/c_\text{s}^2 dr}.
\end{equation}
The plane-wave assumption involves taking all background quantities to be constant and letting
\begin{equation}
    \bar{p} = \hat{p} e^{ikr}, \qquad r^2 \xi^r = \hat{\xi} e^{ikr}.
\end{equation}
This leads to 
\begin{equation}
    \left[ ik - \frac{\mathcal{N}^2}{g} \right] \hat{p}  = (\omega^2- \mathcal{N}^2) \frac{\rho}{r^2} \hat{\xi} e^{\int g/c_\text{s}^2 dr} 
\end{equation}
and
\begin{equation}
    \left[ \frac{ik}{r^2} + \frac{\mathcal{N}^2}{g r^2} \right] \rho \hat{\xi} = \left[ \frac{\lambda(q)}{r^2 \omega^2} - \frac{1}{c_\text{s}^2} + \frac{\mathcal{N}^2}{g^2} \right] \hat{p} e^{-\int g/c_\text{s}^2 dr}.
\end{equation}
This way we arrive at the dispersion relation
\begin{equation}
    \left[ ik - \frac{\mathcal{N}^2}{g} \right] \left[ ik + \frac{\mathcal{N}^2}{g} \right] = - \left( k^2 + \frac{\mathcal{N}^4}{g^2} \right) = (\omega^2- \mathcal{N}^2) \left[  \frac{\lambda(q)}{r^2\omega^2} - \frac{1}{c_\text{s}^2} + \frac{\mathcal{N}^2}{g^2} \right].
\end{equation}
This looks fairly familiar and leads to the usual propagation diagram argument (see, \textit{e.g.}, Refs.~\cite{1989nos..book.....U,2019gwa..book.....A}).
In the barotropic limit ($\mathcal{N}^2=0$), we are left with
\begin{equation}
    - k^2 = \omega^2 \left[ \frac{\lambda(q)}{r^2 \omega^2} - \frac{1}{c_\text{s}^2} \right]
\end{equation}
or 
\begin{equation}
\omega^2 = c_\text{s}^2 \left[ \frac{\lambda(q)}{r^2} + k^2 \right].
\end{equation}
These are the sound waves, which will differ from the usual result as $\lambda$ depends on $\omega$. In the non-rotating limit, we get back the expected Lamb frequency since $\lambda \to l(l+1)$ as $q\to 0$.

In the stratified case ($\mathcal{N}^2 \neq 0$), we get a quadratic for $\omega^2$ (ignoring for the moment the fact that $\lambda$ depends on $q$):
\begin{equation}
    \left( 1 - \frac{\mathcal{N}^2 c_\text{s}^2}{g^2} \right) \omega^4 - \left[ \frac{\lambda(q) c_\text{s}^2}{r^2} + k^2 c_\text{s}^2 + \mathcal{N}^2 \right] \omega^2 + \frac{\lambda(q) \mathcal{N}^2 c_\text{s}^2}{r^2}= 0.
\end{equation}
Here it makes sense to assume that 
\begin{equation}
    \frac{\mathcal{N}^2 c_\text{s}^2}{g^2} \ll 1,
\end{equation}
in which case the roots are
\begin{equation}
    \omega^2 = \frac{1}{2} \left[ \frac{\lambda(q) c_\text{s}^2}{r^2} + k^2 c_\text{s}^2 + \mathcal{N}^2 \right] \left\{ 1 \pm \left[ 1 - \left( \frac{\lambda c_\text{s}^2}{r^2} + k^2 c_\text{s}^2 + \mathcal{N}^2 \right)^{-2}  \frac{4 \lambda \mathcal{N}^2  c_\text{s}^2}{r^2}\right]^{1/2}\right\}.
\end{equation}
Assuming we can Taylor expand the square-root this leads to the two roots
\begin{equation}
    \omega^2 = \left[ \frac{\lambda(q) c_\text{s}^2}{r^2} + k^2 c_\text{s}^2 + \mathcal{N}^2 \right] \left[ 1 - \left( \frac{\lambda c_\text{s}^2}{r^2} + k^2 c_\text{s}^2 + \mathcal{N}^2 \right)^{-2}  \frac{\lambda \mathcal{N}^2 c_\text{s}^2}{r^2}\right]
\end{equation}
and
\begin{equation}
    \omega^2 = \left( \frac{\lambda c_\text{s}^2}{r^2} + k^2 c_\text{s}^2 + \mathcal{N}^2 \right)^{-1} \frac{\lambda \mathcal{N}^2 c_\text{s}^2}{r^2} \approx \left( 1 + \frac{k^2 r^2}{\lambda} \right)^{-1} \mathcal{N}^2.
\end{equation}
This second solution represents the \textit{g}-modes in a stratified star.

Now let us ask a key question: How does the problem change if we insist on the approximation being ``consistent''? Imposing the condition from  Eq.~\eqref{eq:ineq2}, we arrive at the dispersion relation
\begin{equation}
    - \left( k^2 + \frac{\mathcal{N}^4}{g^2} \right) = - \mathcal{N}^2 \left[ \frac{\lambda}{r^2 \omega^2} - \frac{1}{c_\text{s}^2} + \frac{\mathcal{N}^2}{g^2} \right],
 \end{equation}
where we need to keep in mind that we have assumed $\mathcal{N}^2 \gg \omega^2$. This leads to 
\begin{equation}
    \left[ \frac{\mathcal{N}^2}{c_\text{s}^2} + k^2 \right] \frac{r^2}{\lambda} \omega^2 = \mathcal{N}^2.
\end{equation}
Evidently, we have now filtered out the sound waves. In effect, this is a ``sound-proof'' model, which is consistent for low-frequency modes.

\bibliography{bibliography}

\end{document}